\documentclass[prd,amsfonts,twocolumn,nofootinbib,showpacs]{revtex4}
\RequirePackage{graphicx}
\usepackage{enumerate}

\begin{document}

\title{Non-Gaussianity in the inflating curvaton}

\author{Seishi Enomoto}
\affiliation{Department of Physics, Nagoya University, Nagoya 464-8602,
Japan}
\author{Kazunori Kohri}
\affiliation{Cosmophysics group, Theory Center, IPNS, KEK,
and The Graduate University for Advanced Study (Sokendai),
Tsukuba 305-0801, Japan}
\author{Tomohiro Matsuda}
\affiliation{Department of Physics, Lancaster University,  
Lancaster LA1 4YB, UK, and
 Laboratory of Physics, Saitama Institute of Technology,
Fukaya, Saitama 369-0293, Japan}

\begin{abstract}
Inflating curvaton can create curvature perturbation when the curvaton
 density is slowly varying.
Using the delta-N formalism, we discuss the evolution
 of the curvature perturbation during curvaton inflation and find
 analytic formulation of the 
 non-Gaussianity parameter. 
We first consider the inflating curvaton with
 sufficiently long inflationary expansion.
Then we compare the result with short curvaton inflation.
\end{abstract}

\maketitle

\section{Introduction}

The primordial curvature perturbation $\zeta(k)$ that exists  
on cosmological scales just before they start to enter the horizon
is usually related to the perturbations generated during inflation.
Recent observation suggests that $\zeta(k)$ is strongly constrained  and
provides a window on the very early universe~\cite{Lyth-book}. 

The mechanism of generating $\zeta$ can be diverse, but begins presumably
during inflation. 
There are many proposals for generating the curvature perturbation from
the field perturbations, which use one or more fields for the mechanism.

The paradigm of the multi-field inflation scenario has been widely
investigated, but it has usually been supposed that $\zeta(x,t)$
evaluated at an epoch $t_\mathrm{end}$ just after the end of inflation is to be
identified with the observed quantity.  
In this respect, a lot of papers consider the calculation
of the spectrum of $\zeta$ just at the end of
inflation~\cite{Multi1,Multi2,Multi3,Multi4,Multi5,Multi6,
Multi-matsuda, Multi-NG}.

On the other hand, we know that multi-field inflation may
lead to multi-component Universe, in which the curvature
perturbation may evolve significantly. 
The typical example is the curvaton mechanism, in which 
the mixed state of the radiation and the matter causes significant
evolution of the
curvature perturbation~\cite{curvaton-paper, pre-curv, enomoto-matsuda}.
Recently, the curvaton mechanism has been extended to 
include the slowly varying component~\cite{infla-curv, PBH-infcurv}.
The idea of the inflating curvaton has been mentioned earlier in many
papers~\cite{infcurv-pre}.\footnote{Here, we should remember that the
conventional quadratic potential 
(without additional vacuum energy) is
usually not suitable for the inflating curvaton mechanism 
in the sense that the second inflation (curvaton inflation) may not
create enough perturbation via the curvaton mechanism.}
There has been numerical calculation for the hilltop curvaton,
in which the curvaton inflation may take place~\cite{numerical-cal}.
We hope that the reader will be able to compare our analytic calculation
with these preceding works.

Let us first summarize the motivations of this paper.
In the conventional curvaton model there will be a significant
non-Gaussianity when the curvaton decays before it dominates the energy
density.
In that way one usually has the formula $f_{NL}\sim r_\sigma^{-1}$, where 
``$r_\sigma$'' is comparable to the density ratio.
One might suspect that similar mechanism does not work in the inflating
curvaton scenario, simply because the inflating curvaton is already
dominating the density.
This is an important problem.
We must examine the details of the mechanism to answer that question. 
We are basically using the non-linear formalism.
The non-linear formalism is reviewed in the appendix.

If one is going to apply the non-linear formalism into the inflating
 curvaton, one must be careful about the end boundary of the curvaton
mechanisms.
The usual non-linear calculation assumes uniform density hypersurfaces
at the end, where the oscillating curvaton decays.
For the inflating curvaton, the end boundary is where the inflation
ends.
{\it The important point is that the end of inflation
may not coincide with the uniform density hypersurfaces
even though there is no other light scalar field (moduli).
The mismatch (modulation) is due to the conventional isocurvature
 perturbations of the curvaton field, not from perturbations of
 extra light field~\cite{infla-curv, Modulatedcurvatons, Modulatedcurvatons2}.} 
Although the isocurvature perturbations of the curvaton field dies way (or
 convertd into curvature perturbations) during
 the curvaton inflation, it could not be negligible when the curvaton
 inflation is short.
Therefore, after establishing the non-linear formalism of the inflating
 curvaton, we have to go further to deal with the modulation at the
 end.\footnote{Typical examples of the 
 usual modulation will be the modulated reheating 
~\cite{Modulatedcurvatons2, IR, IR-2}, modulated end of inflation~\cite{end-of-inf} or the
inhomogeneous phase transition~\cite{Modulatedcurvatons2, IPhaa}. 
We are not arguing the modulation of this kind, in which additional
light scalar field causes modulation.}
{\it Here the modulation is defined as the 
discrepancy between the transition hypersurfaces and the uniform density
 hypersurfaces.}\footnote{Usually, the curvaton mechanism  assumes
 sudden-decay approximation. 
We are assuming abrupt change of the density scaling relation 
at the ``transition''. 
In the usual curvaton mechanism, modulation at the beginning of the
 curvaton mechanism has been calculated as the deviation in the 
function $g(\sigma_*)$,
which has been introduced by D.H.Lyth in Ref.\cite{Lyth-gfc}. }

We basically use the $\delta N$ formalism for the calculation.
In this formalism, $\zeta$ is defined by smoothing the energy density $\rho$ 
on a super-horizon scale shorter than any scale of interest. 
Then the local energy continuity equation is given by
\begin{equation}
\frac{\partial \rho(x,t) }{\partial t} = - \frac{3}{a(x,t)}
\frac{\partial a(x,t)}
{\partial t} \left( \rho(x,t) + p(x,t) \right),
\end{equation}
where $t$ is time along a comoving thread of spacetime and $a$ is the
scale factor. 

During inflation, the vacuum fluctuation of a light scalar field $\phi_i$ 
is converted at horizon exit to a nearly Gaussian classical perturbation
with spectrum $H/2\pi$.
Here $H\equiv \dot a(t)/a(t)$ is the Hubble parameter defined in the
unperturbed universe.
The $\delta N$  formalism gives
\begin{equation}
\zeta = \delta [ \ln (a(x,t)/a(t_*)] \equiv \delta N, 
\end{equation}
where taking $t_*$ to be an epoch during inflation after relevant scales
leave the horizon, one can assume
$N(\phi_1(x,t_*),\phi_2(x,t_*),\cdots,t,t_*)$ so that  
\begin{equation}
\zeta(x,t) = N_i \delta \phi_i(x,t_*)
+ \frac{1}{2} N_{ij} \delta \phi_i(x,t_*)\delta \phi_j(x,t_*) + \cdots, 
\end{equation}
where a subscript $i$ denotes the derivative with respect to $\phi_i$,
which is evaluated on the unperturbed trajectory.

In the curvaton calculation~\cite{curvaton-paper} one usually
assumes that these expressions are dominated by the single ``curvaton'' field
$\sigma$, which starts oscillation in the radiation dominated Universe
when $\sigma$ has the negligible contribution to the curvature perturbation. 
Then the non-Gaussianity parameter is given by~\cite{Lyth-gfc, SVW}
\begin{eqnarray}
\label{fnl-0}
f_{NL} &\simeq&  \frac{5}{4r_\sigma} \left( 1 +\frac{g''g}{g^{'2}}\right)
-\frac{5}{3} - \frac{5}{6}r_\sigma,
\end{eqnarray}
where $g(\sigma_*)$ is the initial amplitude of the oscillation as a function
of the curvaton field at horizon exit\cite{Lyth-gfc}, and $r_\sigma$
will be defined later in this paper.
The above result is obtained for the two-component Universe, in which
one component behaves like matter while the other behaves like
radiation.
{\em Our calculation is aimed to generalize and extend the above result
to include a slowly varying component.}
Any kind of components (e.g, cosmological
defects~\cite{Topological-curv}) can be included in the same way.

\subsection{Non-linear formalism (brief review)}
Let us first review the basic idea of the non-linear formalism.
(See also the appendix.)
In this paper we consider the non-linear formalism defined in
Ref.~\cite{Lyth-general, Langlois:2008vk}; 
\begin{eqnarray}
\label{def-compzeta}
\zeta_i&=&\delta N+\int^{\rho_i}_{\bar{\rho}_i}
\frac{d\tilde{\rho}_i}{3(1+w_i)\tilde{\rho}_i}\nonumber\\
&=& \label{ln-rho}\delta N + \frac{1}{3(1+w_i)}\ln
 \left(\frac{\rho_i}{\bar{\rho}_i}\right)\nonumber\\
&\simeq & \delta N+\frac{1}{3(1+w_i)}\frac{\delta \rho_i^\mathrm{iso}}{\bar{\rho}_i},
\end{eqnarray}
where $\delta \rho_i^\mathrm{iso}$ will be defined in Eq.(\ref{iso-def}).
Here $w_i=1/3$ for the radiation fluid  and $w_i=0$ for the matter
fluid.
A bar is for the homogeneous quantity, and $\rho_i$ is defined on
the uniform density hypersurfaces.
Due to the isocurvature perturbations, $\bar{\rho}_i(t) \ne \rho_i(x,t)$ is
possible  in the multi-component Universe. 
The curvature perturbation of the total fluid should be
discriminated from the component curvature perturbation $\zeta_i$.
The standard definition of the adiabatic perturbation is given by
\begin{equation}
\delta N=-H \frac{\delta\rho^\mathrm{adi}}{\dot{\rho}},
\end{equation}
where $\delta \rho^\mathrm{adi}\equiv \sum_i\delta \rho_i^\mathrm{adi}$
must be evaluated on the spatially flat hypersurfaces. 
In contrast to $\delta \rho^\mathrm{adi}_i$,
the isocurvature quantity $\delta \rho_i^\mathrm{iso}$ is related to the
fraction perturbation defined on the
uniform density hypersurfaces.
Using the homogeneous density $\bar{\rho}_i(t)$, the isocurvature
density perturbation is defined on the uniform density hypersurfaces as
\begin{equation}
\label{iso-def}
\delta \rho_i^\mathrm{iso}(x,t)\equiv\rho_i(x,t)-\bar{\rho}_i(t),
\end{equation}
which satisfies $\sum_i \delta \rho_i^\mathrm{iso}=0$.
In the multi-component Universe, the hypersurface
defining uniform ``density'' ($\rho\equiv \sum_i\rho_i$) is
 usually different from the one defining uniform ``component
 density'' ($\rho_i$).

We find from the second line of Eq.(\ref{ln-rho}); 
\begin{eqnarray}
\rho_i&=&\bar{\rho}_ie^{3(1+w_i)(\zeta_i-\delta N)}.
\end{eqnarray}
Using the above equation for the two-component Universe, 
the definition of the total energy density 
$\rho^\mathrm{total}\equiv\rho_1+\rho_2=\bar{\rho}_1+\bar{\rho}_2$ on the
uniform density hypersurfaces leads to
\begin{equation}
\label{trivial-iso}
f_1e^{3(1+w_1)(\zeta_1-\delta N)}
+\left(1-f_1 \right)e^{3(1+w_2)(\zeta_2-\delta N)}=1,
\end{equation}
where the fraction of the energy density is defined by
\begin{equation}
f_1 \equiv \frac{\bar{\rho}_1}{\bar{\rho}_1+\bar{\rho}_2}.
\end{equation}
Expanding Eq.(\ref{trivial-iso}) and solving the equation for 
$\delta N$,  we find at first order
\begin{eqnarray}
\label{deltaN-1}
\delta N&=&r_1 \zeta_1+(1-r_1)\zeta_2\nonumber\\
&=& \left[r_1\zeta_1^\mathrm{iso}+(1-r_1)\zeta_2^\mathrm{iso}\right]
+\zeta^\mathrm{adi},
\end{eqnarray}
where $\zeta_i^\mathrm{iso}\equiv \zeta_i-\zeta^\mathrm{adi}$ is
introduced in the last line.
$r_1$ is defined by
\begin{equation}
\label{r-basic}
r_1\equiv
\frac{3(1+w_1)\bar{\rho}_1}{3(1+w_1)\bar{\rho}_1+3(1+w_2)\bar{\rho}_2},
\end{equation}
where $w_1=0$ and $w_2=1/3$ gives for the usual curvaton;
\begin{equation}
r_1\equiv
\frac{3\bar{\rho}_1}{3\bar{\rho}_1+4\bar{\rho}_2}.
\end{equation}
Note that $r_1$ and $f_1$ are comparable in the usual curvaton scenario;
however in the inflating curvaton the discrepancy between these
quantities is significant.
For the slowly varying curvaton density, we consider $1+w_1 \ll 1$.
In that case, Eq.(\ref{r-basic}) can give $r_1\ll 1$ even if the curvaton
density ($\rho_1$) is already dominating the Universe.

Before discussing the evolution of the curvature perturbation for the
slowly varying curvaton density, 
let us remember why the ``evolution'' is
possible in the curvaton mechanism.
Note first that $\zeta^\mathrm{adi}$ is
identical to $\delta N$ at the time when 
the initial quantities are evaluated.
Usual curvaton scenario assumes $\delta N=\zeta^\mathrm{adi}=0$
at the beginning of the oscillation and it defines the initial condition.
Since $w_1=0$ and $w_2=1/3$ are constant in the usual curvaton mechanism,
$\zeta_1$ and $\zeta_2$ are constant during the
evolution~\cite{Lyth-general}.\footnote{This is the reason why the
non-linear formulation is useful for the curvaton
mechanism. Alternatively, the constancy of the scaling exponents can be
used.}
Hence, $\delta N$ is time-dependent only when $r_1(t)$ is changing, which
defines the evolution of the curvature perturbation
in the usual curvaton mechanism.
The definition given by Eq.(\ref{deltaN-1}) can be used anytime and for
any long-wavelength perturbations.
However, because we are formulating the evolution of the adiabatic
perturbation that is caused by the adiabatic-isocurvature mixings,
we need first to define the ``starting point'' at an
epoch and then we can discuss the ``evolution'' thereafter.
In this paper, the quantities evaluated at the starting point are denoted
by the subscript ``ini''.
In that way, at the ``starting point'', $\delta N$ in Eq.(\ref{deltaN-1}) is
identical to $\zeta^\mathrm{adi}_\mathrm{ini}$, however the
evolution causes $\delta N\ne \zeta^\mathrm{adi}_\mathrm{ini}$ 
apart from the starting point.
This is the most basic idea of the curvaton mechanism, which is also
reviewed in the appendix.

When we calculate the evolution of the curvature perturbation using the
curvaton mechanism, $\zeta_{i}$ is evaluated
at the ``starting point'', and $\zeta_i$ is supposed to be constant
thereafter. 
This is the basic requirement for the practical analytic calculation.
Otherwise, numerical
calculation will be mandatory.\footnote{There are many papers in which
numerical calculation has been used for the curvaton mechanism.
See for instance Ref.~\cite{numerical-cal}.}

Besides the ``initial'' quantities that are needed to define 
the curvaton mechanism, the ``primordial'' quantities are
usually evaluated at the end of the primordial inflation (at $t=t_\mathrm{end}$).
Here the corresponding scales exit horizon at $t_*$
($t_*<t_\mathrm{end}\le t_\mathrm{ini}$). 
The initial ($t_\mathrm{ini}$) and the primordial ($t_\mathrm{end}$)
quantities are not necessarily identical.

In this paper we consider the secondary inflation for the
curvaton mechanism (the inflating curvaton). 
The end of
the secondary inflation is denoted by the subscript ``e'', which should
be discriminated from the subscript ``end'' used above.
Usually, the discrepancy between the quantities evaluated at $t_*$ and
$t_\mathrm{ini}$ is explained by the function $g(\phi_{1*})$, which will
be defined later in Eq.(\ref{defg}).

The slowly varying curvaton density 
is realized by $1+w_1\ll 1$ in the above formalism. 
{\it The perturbation of the function 
$\epsilon_w (\phi_1)\equiv (1+w_1)$ is essential.
 This contribution is important, since it
can change the sign of $f_{NL}$.}

Besides the curvaton mechanism, which we have discussed above,
there could be some important contributions from the boundaries.
In the non-linear formalism the evolution ``before'' the starting point is
totally included in the function $g$.
On the other hand, the transition ``at the end of the curvaton mechanism''
 is usually supposed to occur on the uniform density hypersurfaces;
however the inflating curvaton {\it may break the rule}.
In that sense, these two sources from the boundaries must be unified in the 
formalism~\cite{Modulatedcurvatons2}.

\section{Models}
\subsection{Inflating Curvaton 1 (Slow-roll and sufficient inflation)}
\label{pure-inf-curv}

{\bf In this paper  $\sigma$ is the curvaton and $\phi$ is the
inflaton of the primordial inflation.}
The subscripts ``$\sigma$'' and ``$\phi$'' are used to define the related
quantities.
About the densities, $\rho_\sigma$ is the curvaton density and $\rho_r$ is
the radiation density that is generated after the primordial inflation.
For simplicity, we assume that $\phi$ decays directly into radiation
 just after inflation, and $\sigma$ decays into radiation just after the
 curvaton inflation. (The direct-decay approximations.) 
We assume no mixing between $\phi$ and $\sigma$.
$H_I$ denotes the Hubble parameter during the primordial inflation. 
In the direct-decay approximation, the sinusoidal $\sigma$-oscillation
of the curvaton field starts at $H_\mathrm{osc}$ and then it decays
instantly at $H_{\mathrm{dec}}\simeq H_\mathrm{osc}$. 

\begin{figure}[t]
\centering
\includegraphics[width=1.0\columnwidth]{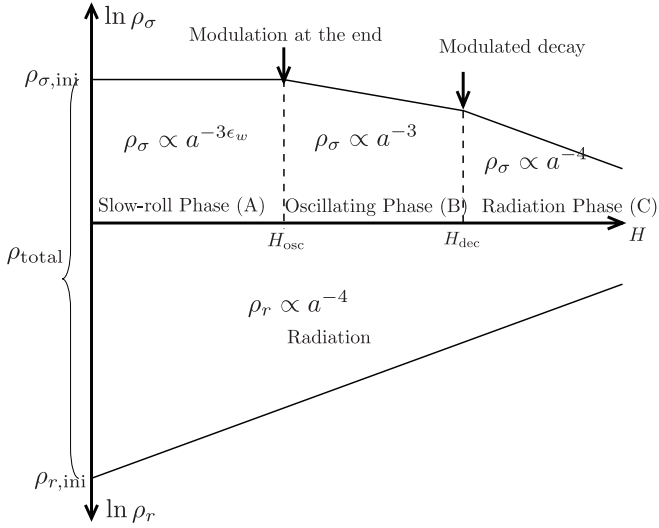}
 \caption{Evolution of the densities $\rho_\sigma$ and $\rho_r$ are
 plotted for the phases (A, B, C). Typical modulation scenarios are shown in the
 figure.
There will be no oscillating phase in the direct-decay limit
 ($H_\mathrm{osc}\simeq H_\mathrm{dec}$).}
\label{fig:density}
\end{figure}

Then, there are two phases $(A,C)$ characterized by the parameter
$w_\sigma$, which are separated by $H=H_\mathrm{osc}\simeq H_\mathrm{dec}$;
\begin{enumerate}[(A)]
\item $\rho_\sigma$ is slowly varying,\\ 
$\rho_r$ is the radiation,\\
 $1+ w_\sigma\equiv \epsilon_w \ll 1$ and $w_r=1/3$.
\item  {\bf This phase is skipped in the direct-decay approximation.}\\
$\sigma$ is oscillating, \\
$\rho_r$ is the radiation,\\
 $w_\sigma = 0$ and $w_r =1/3$
\item Both $\rho_\sigma$ and $\rho_r$ are the radiation, \\
  ($w_\sigma = w_r =1/3$).
\end{enumerate}
In contrast to $\zeta_\sigma$,  $\zeta_r$ is
always invariant during the evolution. (See also Fig.\ref{fig:density})

The evolution of the curvature perturbations in the
phase (A) may not be simple.
We thus need further simplifications of the scenario;
\begin{enumerate}
\item  As far as $\rho_r$ is dominating the Universe, we always find
$r_\sigma \sim \epsilon_w \frac{\rho_\sigma}{\rho_r}<\epsilon_w \ll 1$.
Our first assumption for the analytic calculation is that the curvaton mechanism
       is negligible before the curvaton inflation.

\item Our second assumption is that $\zeta_\sigma$ can behave like
      constant from $t=t_\mathrm{ini}$, somewhat after the beginning of the
      curvaton inflation, and can remain  approximately a constant when the
      significant conversion of 
      the curvaton mechanism are working~\cite{infla-curv}. 
\end{enumerate}
At the end, one will find $r_\sigma\sim 1$
when the curvaton inflation is ``sufficient''; while ``short'' inflation
may lead to $r_\sigma\ll 1$. 
In this section we are considering the former scenario.

At the beginning of the curvaton inflation, we define the
initial perturbation $\zeta_\mathrm{\sigma,ini}$.
Using the same argument as in Eq.(\ref{deltaN-1}), we find just before
the end of the curvaton inflation; 
\begin{eqnarray}
\label{deltanab}
\delta N_- &\equiv& r_{\sigma-} \zeta_{\sigma-} + (1-r_{\sigma-})\zeta_{r-}\nonumber\\
&\simeq& r_{\sigma-}\zeta_{\sigma,\mathrm{ini}}
 +(1-r_{\sigma-})\zeta_{r,\mathrm{ini}}.
\end{eqnarray}
where 
\begin{eqnarray}
\label{r1-inA}
r_{\sigma-} &\equiv&\frac{3\epsilon_{w}\bar{\rho}_{\sigma-}}{3\epsilon_{w}\bar{\rho}_{\sigma-}
+4\bar{\rho}_{r-}}.
\end{eqnarray}
Here the minus sign denotes the quantities just before the end of the
curvaton inflation.
These quantities satisfy
$\bar{\rho}_{\sigma-}\simeq \bar{\rho}_{\sigma,\mathrm{ini}}$ (because $\rho_\sigma$ is
slowly varying) and
$\bar{\rho}_{\sigma,\mathrm{ini}}\simeq \bar{\rho}_{r,\mathrm{ini}}
\gg \bar{\rho}_{r-}$. 

Then, we find the curvature perturbation created by the
evolution~\cite{infla-curv}~\footnote{The standard assumptions of the 
curvaton mechanism are $\zeta_r\ll \zeta_{\sigma}$ 
and $\zeta^\mathrm{adi}_\mathrm{ini}\simeq 0 $.
The above result is obtained using these assumptions.}
\begin{eqnarray}
\label{curvinffin}
\delta N^\mathrm{long}&\simeq &
 r_{\sigma-}\zeta_\mathrm{\sigma,ini}\nonumber\\
&\simeq&
\left[\frac{\delta \rho_{\sigma}}
{3\epsilon_{w}\bar{\rho}_{\sigma}}\right]_\mathrm{ini},
\end{eqnarray}
where $r_{\sigma-}\sim 1$ is assumed for the sufficient inflation.
Modulation at the end is negligible in this model.
(See also Section \ref{hybrid-sec}.)

For the curvaton inflation, we find that the number of e-foldings
required for $r_{\sigma}\simeq 0.9$ (we consider $0.9$ just for instance)
 is given by
\begin{eqnarray}
\label{infcurvn}
\tilde{N}_e&=& \ln
\left(\frac{\bar{\rho}_{\sigma-}}{\bar{\rho}_{r-}}\right)^{1/4}\nonumber\\
&=& -\frac{1}{4}\left[ \ln \epsilon_{w} +
		 \ln\left(\frac{1-r_{\sigma-}}{r_{\sigma-}}\right)
-\ln\frac{4}{3}\right]\nonumber\\
&\simeq & -\frac{1}{4}\left[ \ln \epsilon_{w} +
		 \ln\left(\frac{1}{12}\right)\right],
\end{eqnarray}
where $\tilde{N}_e$ is the number of e-foldings elapsed during the
curvaton inflation.
When $\epsilon_w \ll 1$, it is obvious that significant inflation is
needed to achieve $r_\sigma \sim 1$ before the end.

\subsection{Inflating curvaton 2 (Slow-roll but not sufficient inflation)} 

Practically this model requires fine-tuning of the initial conditions
since we are considering short inflation in the slow-roll limit.
The benefit of the model is that it gives a quite suggestive consequence
in the well-defined limit.

We find from Eq.(\ref{r1-inA});
\begin{equation}
\bar{\rho}_{\sigma-}=\frac{4r_{\sigma-}}{3\epsilon_{w}(1-r_{\sigma-})}\bar{\rho}_{r-}.
\end{equation}
If the curvaton inflation is short (ends with $r_{\sigma-}\ll 1$), 
we find $3\epsilon_{w-}\bar{\rho}_{\sigma-}+4\bar{\rho}_{r-}\simeq
4\bar{\rho}_{r-}$. 
Then, the curvature perturbation created by the
evolution is given by
\begin{eqnarray}
\delta N^\mathrm{short}&\simeq &
 r_{\sigma-}\zeta^\mathrm{iso}_\mathrm{\sigma,ini}\nonumber\\
&\simeq& 
\frac{1}{4}\frac{\delta
\rho_{\sigma,\mathrm{ini}}^\mathrm{iso}}{\bar{\rho}_{r-}}\nonumber\\
&\simeq&
\frac{r_{\sigma-}}{3\epsilon_w}\frac{\delta
\rho_{\sigma,\mathrm{ini}}^\mathrm{iso}}{\bar{\rho}_{\sigma,\mathrm{ini}}}.
\end{eqnarray}
Therefore, $\delta N^\mathrm{short}\ll \delta N^\mathrm{long}$ is
obvious, which shows that longer inflation (curvaton inflation) is
efficient for the curvaton mechanism.

The above result is valid when the end of the curvaton inflation
coincides with the uniform density hypersurfaces; otherwise
the modulation at the end may not be negligible.
This condition is not obvious and highly model-dependent.
The most obvious counter-example is the hybrid curvaton, which will be 
discussed in Sec.\ref{hybrid-sec}.

\section{Non-Gaussianity}
In this section we are basically following the non-linear calculation in
Ref.\cite{SVW}.
We are extending the calculation to include the slowly varying
curvaton density and the perturbations related to $\epsilon_w(\sigma)$.
Lengthy equations are omitted, since
the calculation itself is quite simple and straight.
More details of the expansions will be shown in the appendix.
Writing down the expansions in the non-linear formalism, one will
automatically find $f_{NL}$.

Generically, one can expand
\begin{equation}
\sigma=\bar{\sigma}+\sum_{k=1}^{\infty}\frac{1}{k!}\delta^{(k)}\sigma,
\end{equation}
where $\delta^{(1)}\sigma$ is a Gaussian random field.
In that way, the primordial perturbation can be expanded as
\begin{equation}
\zeta_\sigma=\zeta^{(1)}_\sigma+\sum_{k=2}^{\infty}\frac{1}{k!}
\zeta_\sigma^\mathrm{(k)},
\end{equation}
where $\zeta^{(1)}_\sigma$ is Gaussian.
Non-linearity parameters are defined for the adiabatic perturbation $\zeta$;
\begin{equation}
\zeta=\zeta^{(1)}+\frac{3}{5}f_{NL}(\zeta^{(1)})^2+\frac{9}{25}g_{NL}
(\zeta^{(1)})^3+.... 
\end{equation}

Assume that the curvaton potential during the curvaton inflation is
given by the quadratic potential;
\begin{equation}
\rho_\sigma=V_0\pm \frac{1}{2}m^2 \sigma^2,
\end{equation}
where the effective mass term may have either positive or negative signs.

Using the Gaussian quantum fluctuations at the horizon exit ($\delta
\sigma_{*}$), we can write~\cite{SVW} 
\begin{equation}
\sigma_{*}=\bar{\sigma}_{*}+\delta\sigma_{*}.
\end{equation}
In that case we write 
\begin{equation}
\label{defg}
\sigma_{\mathrm{ini}}\equiv g(\sigma_{*})
\end{equation}
and expand~\cite{SVW}
\begin{equation}
\label{density-ng}
\sigma_{\mathrm{ini}}=\bar{g}+\sum_{k=1}^{\infty}
 \frac{1}{k!}g^{(k)}\left(\frac{\bar{g}}{g'}\frac{\delta\sigma}
{\bar{\sigma}}
\right)^k,
\end{equation}
where we wrote $g^{(n)}\equiv \partial^n g/\partial \sigma_{*}^n$.
Assuming that $\delta N\ll \zeta_\sigma$ for the starting-point
perturbations\footnote{ $\delta N_\mathrm{ini} < \zeta_{\sigma,\mathrm{ini}}$
 is always needed for the curvaton
mechanism. Otherwise the curvaton perturbation cannot dominate the
spectrum.}, we find
\begin{equation}
\label{non-ln-dens}
\rho_{\sigma,\mathrm{ini}}\simeq\bar{\rho}_{\sigma,\mathrm{ini}}e^{3\epsilon_w\zeta_{\sigma,\mathrm{ini}}}.
\end{equation}
Although $\epsilon_w$ is approximately constant during evolution, one
cannot ignore its perturbations. 
We find for $\epsilon_w\propto \sigma^l$:
\begin{equation}
\epsilon_w=\epsilon_w^{(0)}+l\epsilon_w\frac{\delta
 \sigma}{\bar{\sigma}}+....
\end{equation}
We find $l=2$ for the quadratic hilltop potential.
For simplicity, we will use $\epsilon_w$ instead of $\epsilon_w^{(0)}$
when there is no confusion. 

In the calculation below we will omit the subscript ``ini'' when it is
apparent.
Substituting Eq.(\ref{density-ng}) into Eq.(\ref{non-ln-dens}) we obtain
for the inflating curvaton;
\begin{eqnarray}
e^{3\epsilon_w \zeta_\sigma}&=&
\frac{V_0\pm\frac{1}{2}m^2\left[
\bar{g}+\sum_{k=1}^{\infty}
 \frac{1}{k!}g^{(k)}\left(\frac{\bar{g}}{g'}\frac{\delta\sigma}{\bar{\sigma}}
\right)^k
\right]^2}
{V_0\pm \frac{1}{2}m^2 \bar{g}^2},\nonumber\\
\end{eqnarray}
where $\epsilon_w\ll 1$ is assumed for the
inflating curvaton.
Order by order, we have for the expansion 
$\rho_\sigma=\bar{\rho}_\sigma+\sum_{k=1} \frac{1}{k}\delta^{(k)}\rho_\sigma$;
\begin{eqnarray}
\delta^{(1)}\rho_\sigma&=& m^2 g\delta \sigma\\
\delta^{(2)}\rho_\sigma&=& m^2 \left(1+\frac{gg''}{g^{'2}}\right)
(\delta \sigma)^2
\end{eqnarray}
Defining the
ratio $R\equiv \pm\frac{1}{2}m^2\bar{\sigma}^2/\bar{\rho}_\sigma$, we find for
$\zeta_\sigma\equiv
\zeta_\sigma^{(1)}+\frac{1}{2!}\zeta_\sigma^{(2)}+\frac{1}{3!}\zeta_\sigma^{(3)}+...$; 
\begin{eqnarray}
\zeta_\sigma^{(1)}&\simeq& \frac{2R}{3\epsilon_w}
\frac{\delta \sigma}{\bar{\sigma}}\nonumber\\
\zeta^{(2)}_\sigma&=& \frac{3\epsilon_w}{2R}\left(1-2R+\frac{gg''}{g^{'2}}-4\right)
(\zeta_\sigma^{(1)})^2.
\end{eqnarray}
In the last line, 
$-4$ appears from the expansion of $\epsilon_w$.
In the slowly-varying density phase (A), we find
\begin{equation}
\label{nlforminslow}
f_\sigma e^{3\epsilon_w(\zeta_{\sigma}-\delta N)}+(1-f_\sigma)
e^{4(\zeta_r-\delta N)}=1,
\end{equation}
where $f_\sigma\equiv \frac{\bar{\rho}_\sigma}{\bar{\rho}_\sigma+\bar{\rho}_r}$.
Assuming $\zeta_r\simeq 0$ and expanding the equation, we obtain at
first order
\begin{equation}
\label{nlfirstod}
f_\sigma 3\epsilon_w \left[\zeta_\sigma^{(1)}-\delta N^{(1)} \right]
+(1-f_\sigma)\left[-4\delta N^{(1)}\right]
=0,
\end{equation}
where the expansion is given by $\delta N \equiv
\delta N^{(1)}+\frac{1}{2!}\delta N^{(2)}+\frac{1}{3!}\delta N^{(3)}+...$.
We can solve this equation as
\begin{equation}
\delta N^{(1)}\simeq r_\sigma \zeta_\sigma^{(1)}.
\end{equation}
At second order, we find
\begin{eqnarray}
\label{nlsecondod}
&&-4(1-f_\sigma)\delta N^{(2)}+16(1-f_\sigma)(\delta N^{(1)})^2\nonumber\\
&&+3\epsilon_w f_\sigma \left(\zeta_\sigma^{(2)}-\delta N^{(2)}\right)
+9\epsilon_w^2 f_\sigma \left(\zeta_\sigma^{(1)}-\delta N^{(1)}
\right)^2\nonumber\\
&&+6\epsilon_w^{(1)} f\left(\zeta_\sigma^{(1)}-\delta N^{(1)}\right)=0,
\end{eqnarray}
where the last line includes $\epsilon_w$-expansion.
We thus find that
\begin{eqnarray}
\frac{\delta N^{(2)}}{\left(\delta N^{(1)}\right)^2}
&=&\frac{1}{r_\sigma}
\left[\frac{3\epsilon_w}{2R}\left\{
-4r_\sigma-2R+1
+\frac{gg''}{g'^2}\right\}\right.
\nonumber\\
&&\left.+3(1-r_\sigma)^2 \epsilon_w
+4r_\sigma (1-r_\sigma)
\right].
\end{eqnarray}
The above result does NOT reproduce the standard
curvaton even if one substitutes $\epsilon_w=1$ and $R=1$, since our
result has been
obtained after expanding $\epsilon_w\propto \sigma^{2}$.
More details about the expansion and the explicit relation between the
conventional curvaton mechanism will be discussed in the appendix.

We thus find the non-Gaussianity parameter for the slowly-varying
curvaton density;\footnote{This is partly different from earlier simple
estimations~\cite{infla-curv}.} 
\begin{eqnarray}
\label{fnl-1-expand}
f_{NL}&=&
\frac{1}{r_\sigma}
\frac{5\epsilon_w}{4R}\left(1+\frac{gg''}{g^{'2}}\right)\nonumber\\
&&-5\epsilon_w+\frac{10}{3}
+\left(\frac{5}{2}\epsilon_w-\frac{10}{3}\right)r_\sigma-\frac{5\epsilon_w}{R}.
\end{eqnarray}
{\em In contrast to the standard curvaton, a minus sign ($R<0$) is possible
for the hilltop potential~\cite{hilltop-matsuda}}
For the quadratic potential we find $\epsilon_w/R\sim \frac{2}{3}\eta_\sigma$
in the slow-roll limit, where $\eta_\sigma\equiv m^2/3H^2$ is the
conventional slow-roll parameter during 
the curvaton inflation.

\underline{Fast-roll}

For the fast-roll scenario,
 $\epsilon_w$ will have a different
$\eta$-dependence.
For the fast-roll field one will find 
\begin{equation}
\tilde{c}H\dot{\sigma}\simeq V'
\end{equation}
 with the coefficient defined by
\begin{equation}
\tilde{c}=\frac{3+\sqrt{9-12\eta_\sigma}}{2}.
\end{equation}
We thus find
\begin{equation}
-H\frac{\delta \sigma}{\dot{\sigma}}\simeq
\frac{\tilde{c}}{3\eta_\sigma}\frac{\delta \sigma}{\sigma}.
\end{equation}
Our definition of $\epsilon_w$ ($\dot{\rho}=-3H\epsilon_w\rho$)
can be combined with $\dot{\rho}\simeq V'\dot{\sigma}$ to give 
another expression
\begin{equation}
-H\frac{\delta\sigma}{\dot{\sigma}}\simeq\frac{2}{3}\frac{R}{\epsilon_w}
\frac{\delta\sigma}{\sigma}.
\end{equation}
One may choose either $\tilde{c}$ or $\epsilon_w$ for the calculation.
The choice of the definition cannot cause any discrepancy in the result.

\underline{Estimation of the function $g$}

For the practical estimation of the curvature perturbations
one has to calculate $g$.
$g$ is trivial in the slow-roll limit (since in that limit the
motion is negligible), while for the fast-rolling one
has to calculate the model-dependent evolution before the curvaton
mechanism.\footnote{Just to avoid the complexity of the calculations, 
we have ignored $g$ in $\epsilon_w$; note that $g$-dependence appears
from $\epsilon_w^{(2)}$, which does not appear in the calculation of $f_{NL}$.} 
Quadratic potential is an exception, for which $\delta \sigma$ and
$\sigma$ obey identical equation of motion (which is a linear
differential equation) and their combination $\delta \sigma/\sigma$ 
behaves like constant.
Moreover, with the quadratic assumption we have 
\begin{eqnarray}
g'&=&\frac{d g}{d \sigma_{*}}=\frac{g}{\sigma_{*}}\nonumber\\
g''&=& \frac{-g+g'\sigma_{*}}{\sigma_{*}^2}=0,
\end{eqnarray}
which is the result obtained in Ref.\cite{PBH-infcurv} and 
suggests that there is no $g$-dependence in
$f_{NL}$ when the quadratic assumption is valid.
See Ref.\cite{PBH-infcurv} for more details.

\underline{Spectral index and the slow-roll parameters}

The spectral index
requires $|\eta_\sigma|\ll 1$ during primordial inflation, which must 
be explained in the specific inflationary model that creates the
perturbation $\delta \sigma_{*}$.\footnote{According to 
Ref.~\cite{Lyth:2004nx}, one can expect $|\eta_\sigma| \ll 1$ during the
radiation domination epoch. 
The inflating curvaton without mass protection can cause
significant scale-dependence of the perturbations and it may cause
generation of the primordial black 
holes~\cite{PBH-infcurv}.}

\underline{Isocurvature perturbation?}

In the usual curvaton mechanism, $f_{NL}\gg 1$ may lead to
significant isocurvature perturbation~\cite{iso-curv}.
To avoid the unwanted isocurvature perturbation, the baryon number
asymmetry (BAU) must be created after the curvaton mechanism.
{\it This is not the case in the inflating curvaton scenario}, since the energy
density is already dominated by the curvaton even if $r_\sigma$ is smaller
than unity.
This is a unique character of the inflating scenario of the curvaton mechanism.

\underline{About $g$} 

The evaluation of the function $g$ is very important when
one estimates the non-linearity caused by the evolution of the curvaton
field.
Note first that the method required for the evaluation of $g$ is
compatible among different curvaton scenarios 
(the oscillating and the inflating curvatons).
To begin with, we summarize the past discussions about the
function $g$ in the usual (oscillating) curvaton model.
\begin{enumerate}
\item In the (oscillating) curvaton model, the curvature perturbation is
      determined by the quantity $\frac{\delta \sigma}{\sigma}$.
      Useful point is that when the potential is
      quadratic the above combination behaves like constant
      before the oscillation.
      Hence, one usually finds simple estimation {\bf without} knowing
      the explicit 
      form of the function $g$~\cite{Curvatondynamics}. 
\item The simple calculation does not apply when the potential 
      deviates from the quadratic. A detailed study can be
      found in \cite{Curvatondynamics, Enq-Nu}.
      Quantum corrections are discussed in \cite{Enq-LT}.
\end{enumerate}

In the name of the ``curvaton'', the above points apply to the inflating
curvaton, since the evolution before the curvaton mechanism (i.e, the
method required for the estimation of $g$, which is defined at the
beginning of the curvaton mechanism) is
compatible among those curvaton scenarios.

\subsection{More extension (higher-order potential)}

In the above calculation we have considered a simple quadratic potential.
The quadratic potential is convenient for the analytic estimation of the
function $g$;
however the curvaton potential could be different from the quadratic
one.
Here we will find the basic formalism for the calculation.
Analytic estimation of the function $g$ is quite difficult when
the slow-roll conditions are violated.
We consider the potential given by a polynomial;
\begin{equation}
\rho_\sigma=V_0+ \frac{\lambda}{n}\frac{\sigma^n}{M^{n-4}},
\end{equation}
and $\epsilon_w$ could have non-trivial $\sigma$-dependence.
In that case, assuming the instant and direct decay after the slowly varying
density phase (A),
 we find for the inflating curvaton
\begin{eqnarray}
\bar{\rho}_\sigma&=&V_0+ \frac{\lambda}{n}\frac{\sigma^n}{M^{n-4}}\\
\rho_\sigma&=&V_0+\frac{\lambda}{nM^{n-4}}\left[
\bar{g}+\sum_{k=1}\frac{1}{k!}g^{(k)}
\left(\frac{\bar{g}}{g'}\frac{\delta\sigma}{\bar{\sigma}}\right)^k
\right]^n.\nonumber\\
\end{eqnarray}
Defining the ratio
\begin{equation}
R\equiv \frac{\frac{\lambda}{n}\frac{\bar{g}^n}{M^{n-4}}}{\bar{\rho}_\sigma},
\end{equation}
and assuming $\epsilon_w\propto \sigma^l$,
we find order by order;
\begin{eqnarray}
\zeta^{(1)}_\sigma&=& \frac{nR}{3\epsilon_w}\frac{\delta
 \sigma}{\sigma} \simeq -H \frac{\delta \sigma}{\dot{\sigma}}\\
\frac{\zeta^{(2)}_\sigma}{\left[\zeta_{\sigma}^{(1)} 
\right]^2}&=& \frac{3\epsilon_w}{nR}
\left[n-1-nR+\frac{gg''}{g^{'2}}-2l\right]
\end{eqnarray}
In the slowly varying density phase (A), the non-linear formalism gives
Eq.(\ref{nlforminslow}).
Expanding the equation, we obtain Eq.(\ref{nlfirstod}).
At second order, we find
\begin{eqnarray}
\frac{\delta N^{(2)}}{\left(\delta N^{(1)}\right)^2}
&=&\frac{1}{r_\sigma}
\left[\frac{3\epsilon_w}{nR}\left\{
-2lr_\sigma-nR+
(n-1)+\frac{gg''}{g'^2}\right\}\right.
\nonumber\\
&&\left.+3(1-r_\sigma)^2 \epsilon_w
+4r_\sigma (1-r_\sigma)
\right].
\end{eqnarray}
We thus find the non-Gaussianity parameter
\begin{eqnarray}
\label{fnl-2}
f_{NL}&=&\frac{5}{6}\left[
\frac{3\epsilon_w}{nRr_\sigma}\left\{
-2lr_\sigma-nR+
(n-1)+\frac{gg''}{g'^2}\right\}\right.
\nonumber\\
&&\left.+3 \epsilon_w\left(\frac{1}{r_\sigma}-2+r_\sigma\right)
+4(1-r_\sigma)
\right].
\end{eqnarray}
The standard curvaton corresponds to
$n=2$, $l=0$, $R=1$ and $\epsilon_w=1$, which gives
\begin{eqnarray}
\label{standard-compare}
f_{NL}&=&\frac{5}{6}\left[
\frac{3}{2r_\sigma}\left\{
-2+1+\frac{gg''}{g'^2}\right\}\right.
\nonumber\\
&&\left.+3 \left(\frac{1}{r_\sigma}-2+r_\sigma\right)
+4(1-r_\sigma)
\right]\nonumber\\
&=& \frac{5}{4r_\sigma} \left( 1 +\frac{g''g}{g^{'2}}\right)
-\frac{5}{3} - \frac{5}{6}r_\sigma.
\end{eqnarray}

\section{Modulation at the end of the inflating curvaton}
\label{hybrid-sec}

\subsection{Modulation in the non-linear formalism}
The modulation at the end (at the transition $w_{\sigma-}\rightarrow
w_{\sigma+}$) can be implemented in the curvaton mechanism.
Assuming that the transition occurs at the density $\rho_{x}$, the non-linear
formalism {\it evaluated at the end} can be separated as
\begin{eqnarray}
\zeta_\sigma
&=&
\delta N+
\int^{\rho_x}_{\bar{\rho}_\sigma}H
\frac{d\tilde{\rho}_\sigma}{3(1+w_{\sigma+})\tilde{\rho}_\sigma}
+\int^{\rho_\sigma}_{\rho_x}H
\frac{d\tilde{\rho}_\sigma}{3(1+w_{\sigma-})\tilde{\rho}_\sigma}\nonumber\\
&=& \delta N +
\frac{1}{3(1+w_{\sigma+})}\ln
\left(\frac{\rho_{x}}{\bar{\rho}_\sigma}\right)\nonumber\\
&&+\frac{1}{3(1+w_{\sigma-})}\ln
 \left(\frac{\rho_\sigma}{\rho_{x}}\right)
\end{eqnarray}
Here $\bar{\rho}_\sigma$ 
($\rho_\sigma > \rho_{x} > \bar{\rho}_\sigma$)
is placed after the transition.
Allowing oscillation after curvaton inflation (i.e, phase (B) is not
ignored),  
the above quantities are defining the initial condition for the
usual curvaton mechanism during the phase (B).

First take the limit $\rho_{x}\rightarrow
\rho_\sigma$ (i.e, the transition hypersurface is identical to the uniform
 density hypersurface.).
In that case $\zeta_\sigma$ at the end of the inflating curvaton 
is given by
\begin{eqnarray}
\label{ini-for-B}
\zeta_{\sigma,e}
&=&  \delta N + 
\frac{1}{3}\ln
\left(\frac{\rho_\sigma}{\bar{\rho}_\sigma}\right)_e.
\end{eqnarray}
Eq.(\ref{ini-for-B}) gives the initial condition for the
oscillating curvaton that may work after inflation.
Obviously, $\delta \rho_{\sigma,e}\equiv
\rho_{\sigma,e}-\bar{\rho}_{\sigma,e}$
 is negligible after sufficient inflation,
while it could be significant if the curvaton inflation is short.

With regard to the boundary at the end, the opposite limit
$\rho_{x}\rightarrow \bar{\rho}_\sigma$  
is needed for the hybrid curvaton~\cite{hybcurv}, in which
the waterfall begins presumably 
at $\rho_x=\bar{\rho}_\sigma=V(\sigma_{c})$.
In that case we find (in contrast to the conventional calculation);
\begin{eqnarray}
\zeta_{\sigma,e}
&=&  \delta N + 
\frac{1}{3\epsilon_w}\ln
\left(\frac{\rho_\sigma}{\bar{\rho}_\sigma}\right)_e,
\end{eqnarray}
which is enhanced when $\epsilon_w\ll 1$.
The isocurvature density perturbation ($\delta \rho_{\sigma,e}$) 
is indeed significant when the curvaton inflation is short.
Of course the waterfall stage of the hybrid curvaton
may start {\it without curvaton inflation}.
In that case 
the isocurvature perturbation is obviously significant.

Note that the enhancement at the end is not due to the ``evolution''
(the curvaton mechanism), but caused by the ``modulation''.
Unlike the usual modulation, the above
``modulation'' is simply due to the isocurvature perturbation of the
curvaton field. 
There is no additional moduli that causes extra modulation.

\section{Conclusions and discussions}
In this paper we have calculated the curvature perturbation caused by
 the inflating curvaton mechanism.
To explain the mechanism, a typical two-component Universe has been
 considered, in which one component (curvaton) is slowly varying.
In this paper the modulation at the end of the curvaton mechanism is
 specifically defined as the ``inhomogeneous transition in the
 curvaton sector that does not require additional light field''.
The modulation caused by an additional light field has been discussed in 
Ref.~\cite{IPhaa, 
 Modulatedcurvatons2}.

Generically, $f_{NL}$ derived from the non-linear formalism depends on the
model-dependent function $g$.
In that way the analytic estimation of the non-Gaussianity
parameter is possible only when $g$ is obvious.
In the slow-roll limit, $g$ is trivial and the analytic estimation is
possible.
For the fast-roll curvaton, estimation is possible when the potential
is quadratic.
Otherwise the function $g$ requires more assumptions
or highly model-dependent analyses.
Exact calculation of the model-dependent function $g$ has been
 separated from the present work. 
The relation between the curvaton mechanism and the modulation at the
end is clear in the non-linear formalism.
This is the first exact calculation of the non-Gaussianity created 
by the slowly varying curvaton density.

For our purposes, we took some simple set-ups for the calculation.
We thus need further study, such as
\begin{enumerate}
\item Instant phase transition has been considered in this paper; however the
transition should be more complicated depending on the details of the
model parameters~\cite{IPre}.
In the standard curvaton scenario the sudden-decay approximation gives a
good intuitive derivation, however in practice the curvaton density is
continually decaying into radiation.

\item 
In this paper, we were avoiding the interaction between
      components~\cite{modulated-inflation}.  
If the interaction is significant the dissipation may appear,
 which (in the most extreme case) may lead to warm curvaton
      scenario~\cite{matsuda-warm-curvaton}. 

\item In this paper the second component is always the radiation.
This assumption is useful in capturing the essential of the scenario:
      however the assumption may not be valid in practice.
We need to discuss multi-component Universe that could be a mixture of
the slow-varying density, defects, matter and radiation.

\item The curvaton evolution and the modulation
might be significant after N-flation, while the model
considered in this paper is based on two-component Universe.
We thus need some statistical argument for the evolution of the
perturbations after N-flation~\cite{N-flation-paper}.
\end{enumerate}


\section{Acknowledgment}
TM thanks D.~H.~Lyth  and J.~McDonald for many valuable discussions.
This work is supported in part by Grant-in-Aid for Scientific research
from the Ministry of Education, Sci- ence, Sports, and Culture (MEXT),
Japan, No. 21111006, No. 22244030, and No. 23540327 (K.K.).
S.E. is supported by the Grant-in-Aid for Nagoya University Global COE Program,
"Quest for Fundamental Principles in the Universe: from Particles to the Solar
System and the Cosmos".

\appendix
\section{Non-Linear formalism for the conventional curvaton}
In this section we are going to review the basics of the non-linear
formalism applied to the oscillating curvaton mechanism.
The key point in this calculation is the constancy of the component
perturbations.

First remember the non-linear formalism
\begin{eqnarray}
\zeta_\sigma&=&\delta N + \frac{1}{3}\ln
 \left(\frac{\rho_\sigma}{\bar{\rho}_\sigma}\right),\\
\zeta_r&=&\delta N + \frac{1}{4}\ln
 \left(\frac{\rho_r}{\bar{\rho}_r}\right).
\end{eqnarray}
Here $\delta N$ is the perturbation of $N$ between two hypersurfaces;
one is the flat hypersurface, and the other is usually a uniform density
hypersurface.
Besides $\delta N$, we have to define the other quantities
($\rho_\sigma$, $\rho_r$) and ($\bar{\rho}_\sigma$, $\bar{\rho}_r$).
Following the above definition of $\delta N$, we 
define those quantities on the uniform density hypersurface on which
$\delta N$ is defined. 

\begin{figure}[t]
\centering
\includegraphics[width=1.0\columnwidth]{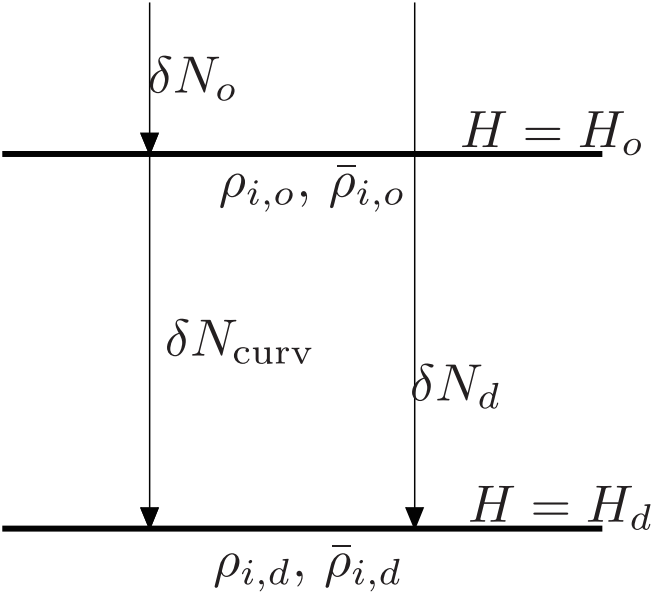}
 \caption{Component perturbations can be evaluated on the uniform density
 hypersurfaces ($H=H_o$ and $H=H_d$). 
Then, using the non-linear formalism one can evaluate $\delta N_o$ and
 $\delta N_d$.
Comparing the results one will obtain $\delta N_\mathrm{curve}\equiv
 \delta N_d -\delta N_o$.
Note that $\rho_{i}(x,t_o)\ne \bar{\rho}_{i}(t_o)$ is possible
on the uniform density hypersurfaces since the Universe is multi-component.}
\label{fig:appendix1}
\end{figure}
We thus find at the uniform density hypersurface ($H=H_o$);
\begin{eqnarray}
\zeta_{\sigma,o}&=&\delta N_o + \frac{1}{3}\ln
 \left(\frac{\rho_{\sigma,o}}{\bar{\rho}_{\sigma,o}}\right),\\
\zeta_{r,o}&=&\delta N_o + \frac{1}{4}\ln
 \left(\frac{\rho_{r,o}}{\bar{\rho}_{r,o}}\right).
\end{eqnarray}
See also Fig.(\ref{fig:appendix1}) for the definition.
Solving these equations we find
\begin{eqnarray}
\label{eq-o}
\rho_{\sigma,o} &=&\bar{\rho}_{\sigma,o}e^{3(\zeta_{\sigma,o}-\delta N_o)},\nonumber\\
\rho_{r,o} &=&\bar{\rho}_{r,o}e^{4(\zeta_{r,o}-\delta N_o)}.
\end{eqnarray}
The trivial identity is
\begin{equation}
\frac{\rho_{\sigma,o}+\rho_{r,o}}{\bar{\rho}_{\sigma,o}+\bar{\rho}_{r,o}}=1,
\end{equation}
where $\rho_{\sigma,o}$ and $\rho_{r,o}$ can be replaced using
Eq.(\ref{eq-o}).
We find the equation
\begin{eqnarray}
\bar{f}_{\sigma,o} e^{3(\zeta_{\sigma,o}-\delta N_o)}+
(1-\bar{f}_{\sigma,o}) e^{4(\zeta_{r,o}-\delta N_o)}&=&1,
\end{eqnarray}
where the ratio is defined by
\begin{eqnarray}
\bar{f}_{\sigma,o}&=&
 \frac{\bar{\rho}_{\sigma,o}}{\bar{\rho}_{\sigma,o}+\bar{\rho}_{r,o}}.
\end{eqnarray}
Considering the expansion
\begin{equation}
e^X=1+X+\frac{1}{2}X^2+..,
\end{equation}
we find at first order
\begin{eqnarray}
\delta N_o&=& r_{\sigma,o}\zeta_{\sigma,o}
 +(1-r_{\sigma,o})\zeta_{r,o}\nonumber\\
&=&r_{\sigma,o}\left[\delta N_o + \frac{1}{3}\ln
 \left(\frac{\rho_{\sigma,o}}{\bar{\rho}_{\sigma,o}}\right)\right]\nonumber\\
&& +(1-r_{\sigma,o})\left[\delta N_o + \frac{1}{4}\ln
 \left(\frac{\rho_{r,o}}{\bar{\rho}_{r,o}}\right)\right]\nonumber\\
&=& \delta N_o + \frac{r_o}{3}\ln
 \left(\frac{\rho_{\sigma,o}}{\bar{\rho}_{\sigma,o}}\right)
+\frac{(1-r_o)}{4}\ln
 \left(\frac{\rho_{r,o}}{\bar{\rho}_{r,o}}\right).\nonumber\\
\end{eqnarray}
The trivial identity is
\begin{eqnarray}
&&\frac{r_o}{3}\ln
 \left(\frac{\rho_{\sigma,o}}{\bar{\rho}_{\sigma,o}}\right)
+\frac{(1-r_o)}{4}\ln
 \left(\frac{\rho_{r,o}}{\bar{\rho}_{r,o}}\right)\nonumber\\
&=&\frac{\bar{\rho}_{\sigma,o}
\ln \left(\frac{\rho_{\sigma,o}}{\bar{\rho}_{\sigma,o}}\right)
+\bar{\rho}_{r,o}\ln \left(\frac{\rho_{r,o}}{\bar{\rho}_{r,o}}\right)
}{3\bar{\rho}_{\sigma,o}+4\bar{\rho}_{r,o}}\nonumber\\
&=&0.
\end{eqnarray}
Equivalently, for the expansion $\delta \rho_i\equiv
\rho_i-\bar{\rho}_i$ the above equation gives
\begin{eqnarray}
\delta \rho_{\sigma,o}+\delta \rho_{r,o}&=&0,
\end{eqnarray}
where the trivial expansion used above is
\begin{equation}
\ln(1+X)=X-\frac{1}{2}X^2+...
\end{equation}

{\bf Here, the point 
is that the component perturbations are constant
during the curvaton mechanism.}

One can evaluate the non-linear formalism {\bf away} from
$H=H_o$.
Choosing {\bf another} hypersurface at $H=H_d$, one can evaluate
\begin{eqnarray}
\delta N_d&=& r_{\sigma,d}\zeta_{\sigma,d}
 +(1-r_{\sigma,d})\zeta_{r,d}\nonumber\\
&=& r_{\sigma,d}\zeta_{\sigma,o}
 +(1-r_{\sigma,d})\zeta_{r,o},
\end{eqnarray}
and $\delta N_\mathrm{curv}\equiv 
\delta N_d-\delta N_o$ gives the ``evolution'' between the two
hypersurfaces.
We thus find for $r_{\sigma,d}\gg r_{\sigma,o}$:
\begin{eqnarray}
\delta N_\mathrm{curv}
&=& (r_{\sigma,d}-r_{\sigma,o})\zeta_{\sigma,o}
-(r_{\sigma,d}-r_{\sigma,o})\zeta_{r,o}\nonumber\\
&\simeq&
r_{\sigma,d}\left[\frac{\delta \rho_{\sigma,o}}{3\bar{\rho}_{\sigma,o}}\right].
\end{eqnarray}
Note that $\delta N_o$ disappears from the result because of the obvious
cancellation. 

If one chooses $H_o=H_\mathrm{osc}$ at the beginning of the curvaton
oscillation and $H_d=H_\mathrm{dec}$ at the decay, the above result gives the evolution
of the curvature perturbation in the conventional curvaton mechanism.
If one assumes $\delta N_\mathrm{inf}\simeq \delta N_\mathrm{osc}$, which
explains the curvature perturbation generated during primordial inflation,
one will find 
\begin{eqnarray}
\delta N_\mathrm{dec} &\equiv& \delta N_{\mathrm{inf}}+ \delta N_\mathrm{curv}.
\end{eqnarray}

\section{More about the expansion}
When we calculate the expansion of the component perturbation
$\zeta_\sigma$, we expand the relation $\bar{\rho}_\sigma
e^{3\epsilon_w \zeta_\sigma}=\rho_\sigma$, which gives
\begin{eqnarray}
\label{expand1}
&&\left[V_0+\frac{\lambda}{n}\frac{\bar{g}^n}{M^{n-4}}\right]
e^{3\left(\epsilon_{w}^{(0)}+\epsilon_{w}^{(1)}+\frac{1}{2!}\epsilon_{w}^{(2)}+...\right)
\left[\zeta_{\sigma}^\mathrm{(1)} 
+\frac{1}{2}\zeta_{\sigma}^\mathrm{(2)}
+\frac{1}{6}\zeta_{\sigma}^\mathrm{(3)}+..
\right]}\nonumber\\
&=&
V_0+
\frac{\lambda}{nM^{n-4}}\left[
\bar{g}+\sum_{k=1}\frac{1}{k!}g^{(k)}
\left(\frac{\bar{g}}{g'}\frac{\delta^{(1)}\sigma}{\bar{\sigma}}\right)^k
\right]^n,
\end{eqnarray}
where $\epsilon_w$ is expanded for $\epsilon_w\propto \sigma^l$;
\begin{equation}
\epsilon_w=\epsilon_w^{(0)}+l\epsilon_w\frac{\delta^{(1)}\sigma}{\bar{\sigma}}
+...
\end{equation}
The left-hand side of (\ref{expand1}) has
\begin{eqnarray}
&& e^{3\left(\epsilon_{w}^{(0)}+\epsilon_{w}^{(1)}+\frac{1}{2!}\epsilon_{w}^{(2)}+...\right)
 \left[\zeta_{\sigma}^\mathrm{(1)}
  +\frac{1}{2}\zeta_{\sigma}^\mathrm{(2)} 
+\frac{1}{6}\zeta_{\sigma}^\mathrm{(3)}+..
\right]}\nonumber\\
&=&[1]_\mathrm{0th}
+\left[3\epsilon_{w}^{(0)}\zeta_{\sigma}^\mathrm{(1)}\right]_\mathrm{1st}\nonumber\\
&&+\left[3\epsilon_{w}^{(1)}\zeta_{\sigma}^\mathrm{(1)}
+\frac{3}{2}\epsilon_{w}^{(0)}\zeta_{\sigma}^\mathrm{(2)}
+\frac{1}{2}\left(3\epsilon_{w}^{(0)}\zeta_{\sigma}^\mathrm{(1)}
\right)^2\right]_\mathrm{2nd}\nonumber\\
&&+...,
\end{eqnarray}
where the subscripts 0th, 1st and 2nd are added to show clearly the
expansion. 
The right-hand side of Eq.(\ref{expand1}) has
\begin{eqnarray}
&&\left[
\bar{g}+\sum_{k=1}\frac{1}{k!}g^{(k)}
\left(\frac{\bar{g}}{g'}\frac{\delta^{(1)}\sigma}{\bar{\sigma}}\right)^k
\right]^n\nonumber\\
&=&\bar{g}^n\left[1\right]_\mathrm{0th}
+\bar{g}^n\left[n\frac{\delta\sigma}{\bar{\sigma}}\right]_\mathrm{1st}\nonumber\\
&&+\bar{g}^n\left[\left\{\frac{n(n-1)}{2}
+\frac{n}{2}\frac{gg''}{g'^2}\right\}\left(\frac{\delta \sigma}{\sigma}\right)^2\right]_\mathrm{2nd}.
\end{eqnarray}
Defining
$\bar{\rho}_\sigma\equiv\left[V_0+\frac{\lambda}{n}\frac{\bar{g}^n}{M^{n-4}}\right]$, 
the 0-th order terms on both sides give the relation
\begin{equation}
\bar{\rho}=\bar{\rho},
\end{equation}
which is trivial.

The first order relation is
\begin{equation}
\bar{\rho}_\sigma\times 3\epsilon_{w}^{(0)}\zeta_{\sigma}^\mathrm{(1)} 
= \frac{\lambda\bar{g}^n}{nM^{n-4}}
\left[n\frac{\delta\sigma}{\bar{\sigma}}\right],
\end{equation}
which leads to
\begin{eqnarray}
\label{0th-zeta}
\zeta^{\mathrm{iso}(1)}_\sigma&=& \frac{nR}{3\epsilon_w^{(0)}}\frac{\delta
 \sigma}{\sigma},
\end{eqnarray}
where we have introduced the ratio 
$R\equiv \frac{\frac{\lambda\bar{g}^n}{nM^{n-4}}}{\bar{\rho}_\sigma}$.

The second order equation is
\begin{eqnarray}
&&\left[3\epsilon_{w}^{(1)}\zeta_{\sigma}^\mathrm{(1)}
+\frac{3}{2}\epsilon_{w}^{(0)}\zeta_{\sigma}^\mathrm{(2)}
+\frac{1}{2}\left(3\epsilon_{w}^{(0)}\zeta_{\sigma}^\mathrm{(1)}
\right)^2\right]\nonumber\\
&=&R\left[\frac{n(n-1)}{2}
+\frac{n}{2}\frac{gg''}{g'^2}\right]\left(\frac{\delta \sigma}{\sigma}\right)^2.
\end{eqnarray}
We find for $\epsilon_w^\mathrm{(0)}\propto \sigma^l$;
\begin{eqnarray}
3\epsilon_{w}^{(1)}\zeta_{\sigma}^\mathrm{(1)}
&=&3l\epsilon_w^\mathrm{(0)}\frac{\delta
\sigma}{\sigma}\zeta_{\sigma}^\mathrm{(1)}\nonumber\\
&=&\frac{9l}{n}\frac{\left(\epsilon_w^\mathrm{(0)}\right)^2}{R}
\left(\zeta_{\sigma}^\mathrm{(1)}
\right)^2,
\end{eqnarray}
where the last equation uses (\ref{0th-zeta}).
As far as there will be no confusion in the calculations,
we are going to replace
$\epsilon_w^{(0)}\rightarrow \epsilon_w$ for simplicity .
We find for the second order relation;
\begin{eqnarray}
&&\frac{9l}{n}\frac{\epsilon_w^2}{R}\left(\zeta_{\sigma}^\mathrm{(1)}
\right)^2+\frac{3}{2}\epsilon_{w}\zeta_{\sigma}^\mathrm{(2)}
+\frac{1}{2}\left(3\epsilon_{w}\zeta_{\sigma}^\mathrm{(1)}
\right)^2\nonumber\\
&=&\frac{9\epsilon_w^2}{n^2R}
\left[\frac{n(n-1)}{2}
+\frac{n}{2}\frac{gg''}{g'^2}\right]
\left(\zeta_{\sigma}^{(1)}\right)^2,
\end{eqnarray}
where the last equation uses (\ref{0th-zeta}).
Solving the above equation for $\zeta_\sigma^\mathrm{(2)}$, we find
\begin{eqnarray}
\zeta^{\mathrm{(2)}}_\sigma&=& 
\frac{3\epsilon_w}{nR}\left\{
-2l-nR+\frac{2}{n}
\left[\frac{n(n-1)}{2}
+\frac{n}{2}\frac{gg''}{g'^2}\right]\right\}
\left(\zeta_{\sigma}^{\mathrm{(1)}}\right)^2\nonumber\\
&&
\end{eqnarray}
The standard curvaton scenario corresponds to $\epsilon_w=1$ ($l=0$),
$n=2$ (quadratic potential) and $R=1$ ($V_0=0$), which give
the conventional curvaton result~\cite{SVW} for the component perturbation
\begin{eqnarray}
\zeta^{\mathrm{(2)}}_\sigma&=& 
\frac{3}{2}\left[-1+\frac{gg''}{g'^2}\right]
\left(\zeta_{\sigma}^{\mathrm{(1)}}\right)^2.
\end{eqnarray}
We thus finished the expansion of the component perturbation $\zeta_\sigma$.

In order to calculate the curvature perturbation $\zeta$ from the
component perturbation $\zeta_\sigma$, we use the equation
\begin{equation}
f_\sigma e^{3\epsilon_w(\zeta_{\sigma}-\delta N)}+(1-f_\sigma)
e^{-4\delta N}=1,
\end{equation}
where $\zeta_r$ has been neglected.
Expanding $e^X=1+X+\frac{1}{2}X^2+...,$ for
$X=3\epsilon_w(\zeta_{\sigma}-\delta N)$ and
$X=-4\delta N$, we obtain
\begin{eqnarray}
&&f_\sigma
\left[1+3\left(
\epsilon_{w}^{(0)}+\epsilon_{w}^{(1)}+\frac{1}{2!}\epsilon_{w}^{(2)}+...
\right)\right. \nonumber\\
&& \times\left.\left(
\zeta_\sigma^{(1)}+\frac{1}{2}\zeta_\sigma^{(2)}
-\delta N^{(1)}+\frac{1}{2}\delta N^{(2)}+...\right)\right.\nonumber\\ 
&&+\frac{3^2}{2}\left(
\epsilon_{w}^{(0)}+\epsilon_{w}^{(1)}+\frac{1}{2!}\epsilon_{w}^{(2)}+...
\right)^2\nonumber\\
&&\times\left.\left(
\zeta_\sigma^{(1)}+\frac{1}{2}\zeta_\sigma^{(2)}
-\delta N^{(1)}+\frac{1}{2}\delta N^{(2)}+...\right)^2+...\right]\nonumber\\
&&+(1-f_\sigma)
\left[
1-4\left(\delta N^{(1)}+\frac{1}{2}\delta N^{(2)}+...\right)\right.
\nonumber\\ 
&&\left.+\frac{4^2}{2}\left(\delta N^{(1)}+\frac{1}{2}\delta N^{(2)}+...\right)^2
\right]\nonumber\\
&=&1.
\end{eqnarray}
At 0-th order we find
\begin{equation}
f_\sigma+(1-f_\sigma)=1,
\end{equation}
which is trivial.
At the first order we find
\begin{equation}
3f_\sigma \epsilon_w^{(0)} \left[\zeta_\sigma^{(1)}-\delta N^{(1)} \right]
+(1-f_\sigma)\left[-4\delta N^{(1)}\right]
=0.
\end{equation}
Introducing 
$r_\sigma\equiv
\frac{3\epsilon_w^{(0)}f_\sigma}{4(1-f_\sigma)+3\epsilon_w^{(0)}
f_\sigma}$, 
we can solve this equation as
\begin{equation}
\delta N^{(1)}=r_\sigma \zeta_\sigma^{(1)}.
\end{equation}
At second order, we find
\begin{eqnarray}
&&3 f_\sigma\epsilon_w^{(1)}  
\left(\zeta_\sigma^{(1)}-\delta N^{(1)}\right)+\frac{3}{2} f_\sigma\epsilon_w^{(0)}  
\left(\zeta_\sigma^{(2)}-\delta N^{(2)}\right)\nonumber\\
&&+\frac{9}{2}f_\sigma\left(\epsilon_w^{(0)}\right)^2
\left(\zeta_\sigma^{(1)}-\delta N^{(1)}
\right)^2\nonumber\\
&&-2(1-f_\sigma)\delta N^{(2)}+8(1-f_\sigma)(\delta N^{(1)})^2\nonumber\\
&=&0.
\end{eqnarray}
We have to solve this equation for $\delta N^{(2)}$.
We are going to use the relations
\begin{eqnarray}
\epsilon_w^{(1)}&=& l\epsilon_w \frac{\delta \sigma}{\sigma}
=\frac{3l\epsilon_w^2}{nR}\zeta_\sigma^{(1)}\\
\zeta_\sigma^{(1)}-\delta N^{(1)}&=&(1-r_\sigma)\zeta_\sigma^{(1)}\\
\frac{\zeta^{\mathrm{(2)}}_\sigma}
{\left(\zeta_{\sigma}^{\mathrm{(1)}}\right)^2}&=& 
\frac{3\epsilon_w}{nR}\left\{
-2l-nR+\right.\nonumber\\
&&\left.\frac{2}{n}
\left[\frac{n(n-1)}{2}
+\frac{n}{2}\frac{gg''}{g'^2}\right]\right\}.
\end{eqnarray}
Therefore, the second order equation can be rewritten as
\begin{eqnarray}
&&\frac{1}{2}\left[3f_\sigma\epsilon_w^{(0)}+4(1-f_\sigma)\right]\delta
 N^{(2)}\nonumber\\
&=&\frac{\left(\delta N^{(1)}\right)^2}{r_\sigma^2}
\left[\frac{9 f_\sigma l(\epsilon_w^{(0)})^2}{nR}
(1-r_\sigma)\right.\nonumber\\
&&+ 
\frac{9f_\sigma(\epsilon_w^{(0)})^2}{2nR}\left\{
-2l-nR+\frac{2}{n}
\left[\frac{n(n-1)}{2}
+\frac{n}{2}\frac{gg''}{g'^2}\right]\right\}
\nonumber\\
&&\left.+\frac{9}{2}f_\sigma (1-r_\sigma)^2 \left(\epsilon_w^{(0)}\right)^2
+8r_\sigma^2 (1-f_\sigma)\right].
\end{eqnarray}
Using the equation
\begin{equation}
\frac{1-f_\sigma}{f_\sigma}=\frac{3\epsilon_w^{(0)}}{4r_\sigma}(1-r_\sigma),
\end{equation}
we can write
\begin{eqnarray}
&&\frac{3f_\sigma\epsilon_w}{2r_\sigma}\delta N^{(2)}
\nonumber\\
&=&\frac{\left(\delta N^{(1)}\right)^2}{r_\sigma^2}
\left[\frac{9 f_\sigma l\epsilon_w^2}{nR}
(1-r_\sigma)\right.\nonumber\\
&&+ \frac{9f_\sigma\epsilon_w^2}{2nR}\left\{
-2l-nR +\frac{2}{n}
\left[\frac{n(n-1)}{2}
+\frac{n}{2}\frac{gg''}{g'^2}\right]\right\}
\nonumber\\
&&\left.+\frac{9}{2}f_\sigma (1-r_\sigma)^2 \left(\epsilon_w\right)^2
+6r_\sigma f_\sigma\epsilon_w(1-r_\sigma)
\right]\nonumber\\
&=&\frac{\left(\delta N^{(1)}\right)^2}{r_\sigma^2}\times\nonumber\\
&&\left[\frac{9f_\sigma\epsilon_w^2}{2nR}\left\{
-2lr_\sigma-nR+\frac{2}{n}
\left[\frac{n(n-1)}{2}
+\frac{n}{2}\frac{gg''}{g'^2}\right]\right\}\right.
\nonumber\\
&&\left.+\frac{9}{2}f_\sigma (1-r_\sigma)^2 \left(\epsilon_w\right)^2
+6r_\sigma f_\sigma\epsilon_w(1-r_\sigma)
\right]
\end{eqnarray}
Finally, we find
\begin{eqnarray}
\frac{\delta N^{(2)}}{\left(\delta N^{(1)}\right)^2}
&=&\frac{2}{3f_\sigma\epsilon_wr_\sigma}\times\nonumber\\
&&\left[\frac{9f_\sigma\epsilon_w^2}{2nR}
\left\{
-2lr_\sigma-nR
\left[(n-1)
+\frac{gg''}{g'^2}\right]\right\}\right.
\nonumber\\
&&\left.+\frac{9}{2}f_\sigma (1-r_\sigma)^2 \left(\epsilon_w\right)^2
+6r_\sigma f_\sigma\epsilon_w(1-r_\sigma)
\right]\nonumber\\
&=&\frac{1}{r_\sigma}
\left[\frac{3\epsilon_w}{nR}\left\{
-2lr_\sigma-nR+
(n-1)+\frac{gg''}{g'^2}\right\}\right.
\nonumber\\
&&\left.+3(1-r_\sigma)^2 \epsilon_w
+4r_\sigma (1-r_\sigma)
\right].
\end{eqnarray}
Therefore, the non-linear parameter $f_{NL}$ is
\begin{eqnarray}
f_{NL}&=&\frac{5}{6}\left[
\frac{3\epsilon_w}{nRr_\sigma}\left\{
-2lr_\sigma-nR+
(n-1)+\frac{gg''}{g'^2}\right\}\right.
\nonumber\\
&&\left.+3 \epsilon_w\left(\frac{1}{r_\sigma}-2+r_\sigma\right)
+4(1-r_\sigma)
\right]\nonumber\\
&=&\frac{5}{6}\left[
\frac{3\epsilon_w}{nRr_\sigma}\left\{
-2lr_\sigma+
(n-1)+\frac{gg''}{g'^2}\right\}\right.
\nonumber\\
&&\left.+3 \epsilon_w\left(-2+r_\sigma\right)
+4(1-r_\sigma)
\right].
\end{eqnarray}
The standard curvaton scenario corresponds to $\epsilon_w=1$ ($l=0$),
$n=2$ (quadratic potential) and $R=1$ ($V_0=0$), which give
the conventional curvaton result~\cite{SVW}.

\section{How to compare Eq.(\ref{fnl-1-expand}) and Eq.(\ref{fnl-2})
 with conventional curvaton Eq.(\ref{fnl-0})}

In this section we are going to compare our result of the generalized
curvaton calculation with the standard curvaton result
Eq.(\ref{fnl-0}), to show clearly how one can obtain the conventional
curvaton from the generalized formulation.

Our first calculation gives Eq.(\ref{fnl-1-expand}); however if one
needs to compare the result with Eq.(\ref{fnl-0}), one will soon find
that the equation contains many new parameters that are making the
equation rather confusing, and what is worse, one is required to remove
the contribution coming from the expansion with respect to the parameter
$\epsilon_w$, since in the original curvaton scenario $\epsilon_w$ is a
constant that does not contribute to the expansion.
To remove the expansion of $\epsilon_w$, it would be useful to
use the more generalized formula (\ref{fnl-2}), in which $l=0$
corresponds to a ``constant $\epsilon_w$''.
Then the other paramaters are; $n=2$ gives the quadratic potential for the
curvaton, $R=1$ corresponds to $V_0=0$, and finally $\epsilon_w=1$ means
$\rho_\sigma \propto a^{-3}$.
Substituting those quantities into Eq.(\ref{fnl-2}), one will find
Eq.(\ref{standard-compare}).


\begin{thebibliography}{1}
\bibitem{Lyth-book}
 D.~H.~Lyth, A.~R.~Liddle,
  ``The primordial density perturbation: Cosmology, inflation and the origin of structure,''
  Cambridge, UK: Cambridge Univ. Pr. (2009) 497 p.

\bibitem{Multi1}
  D.~Wands, K.~A.~Malik, D.~H.~Lyth and A.~R.~Liddle,
  ``A New approach to the evolution of cosmological perturbations on large scales,''
  Phys.\ Rev.\ D {\bf 62}, 043527 (2000)
  [astro-ph/0003278].
\bibitem{Multi2}
 A.~Gangui, F.~Lucchin, S.~Matarrese and S.~Mollerach,
  ``The Three point correlation function of the cosmic microwave background in inflationary models,''
  Astrophys.\ J.\  {\bf 430}, 447 (1994)
  [astro-ph/9312033].
\bibitem{Multi3}
 M.~Sasaki and E.~D.~Stewart,
  ``A General analytic formula for the spectral index of the density perturbations produced during inflation,''
  Prog.\ Theor.\ Phys.\  {\bf 95}, 71 (1996)
  [astro-ph/9507001].
\bibitem{Multi4}
  J.~Garcia-Bellido and D.~Wands,
  ``Metric perturbations in two field inflation,''
  Phys.\ Rev.\ D {\bf 53}, 5437 (1996)
  [astro-ph/9511029].
\bibitem{Multi5}
 V.~F.~Mukhanov and P.~J.~Steinhardt,
  ``Density perturbations in multifield inflationary models,''
  Phys.\ Lett.\ B {\bf 422}, 52 (1998)
  [astro-ph/9710038].
\bibitem{Multi6}
 D.~Polarski and A.~A.~Starobinsky,
  ``Isocurvature perturbations in multiple inflationary models,''
  Phys.\ Rev.\ D {\bf 50}, 6123 (1994)
  [astro-ph/9404061].
\bibitem{Multi-matsuda}
  T.~Matsuda,
  ``Delta-N formalism for the evolution of the curvature perturbations in generalized multi-field inflation,''
  Phys.\ Lett.\ B {\bf 682}, 163 (2009)
  [arXiv:0906.2525 [hep-th]];
T.~Matsuda,
``Elliptic Inflation: Generating the curvature perturbation without
	slow-roll,'' 
  JCAP {\bf 0609}, 003 (2006)
  [hep-ph/0606137].
\bibitem{Multi-NG}
 F.~Bernardeau and J.~-P.~Uzan,
  ``NonGaussianity in multifield inflation,''
  Phys.\ Rev.\ D {\bf 66}, 103506 (2002)
  [hep-ph/0207295];
  L.~E.~Allen, S.~Gupta and D.~Wands,
  ``Non-gaussian perturbations from multi-field inflation,''
  JCAP {\bf 0601}, 006 (2006)
  [astro-ph/0509719];
  H.~Assadullahi, H.~Firouzjahi, M.~H.~Namjoo and D.~Wands,
  ``Curvaton and the inhomogeneous end of inflation,''
  arXiv:1207.7006 [astro-ph.CO].



\bibitem{curvaton-paper}
  T.~Moroi, T.~Takahashi,
  ``Effects of cosmological moduli fields on cosmic microwave background,''
  Phys.\ Lett.\  {\bf B522}, 215-221 (2001).
  [hep-ph/0110096].
  D.~H.~Lyth, D.~Wands,
  ``Generating the curvature perturbation without an inflaton,''
  Phys.\ Lett.\  {\bf B524}, 5-14 (2002).
  [hep-ph/0110002].

\bibitem{pre-curv}
  Z.~G.~Berezhiani, A.~S.~Sakharov and M.~Y.~.Khlopov,
  ``Primordial background of cosmological axions,''
  Sov.\ J.\ Nucl.\ Phys.\  {\bf 55}, 1063 (1992)
  [Yad.\ Fiz.\  {\bf 55}, 1918 (1992)];
 A.~S.~Sakharov, D.~D.~Sokoloff and M.~Y.~.Khlopov,
  ``Large scale modulation of the distribution of coherent oscillations of a primordial axion field in the universe,''
  Phys.\ Atom.\ Nucl.\  {\bf 59}, 1005 (1996)
  [Yad.\ Fiz.\  {\bf 59N6}, 1050 (1996)].
\bibitem{enomoto-matsuda}
S.~Enomoto and T.~Matsuda,
``Curvaton mechanism after multi-field inflation'', to appear


\bibitem{infla-curv}
 K.~Dimopoulos, K.~Kohri, D.~H.~Lyth and T.~Matsuda,
  ``The inflating curvaton,''
  JCAP {\bf 1203}, 022 (2012)
  [arXiv:1110.2951 [astro-ph.CO]];
\bibitem{PBH-infcurv}
 K.~Kohri, C.~-M.~Lin and T.~Matsuda,
  ``PBH from the inflating curvaton,''
  arXiv:1211.2371 [hep-ph].

\bibitem{infcurv-pre}
  D.~Polarski and A.~A.~Starobinsky,
intermediate matter dominated stage,''
Nucl.\ Phys.\ B {\bf 385}, 623 (1992);
  D.~Polarski and A.~A.~Starobinsky,
Phys.\ Rev.\ D {\bf 50}, 6123 (1994)  [astro-ph/9404061];
  D.~Langlois and F.~Vernizzi,
Phys.\ Rev.\ D {\bf 70}, 063522 (2004)  [astro-ph/0403258];
  T.~Moroi, T.~Takahashi and Y.~Toyoda,
Phys.\ Rev.\ D {\bf 72}, 023502 (2005)  [hep-ph/0501007];
  K.~Ichikawa, T.~Suyama, T.~Takahashi and M.~Yamaguchi,
Inflaton and Curvaton Models,''
 Phys.\ Rev.\ D {\bf 78}, 023513 (2008)  [arXiv:0802.4138 [astro-ph]];

\bibitem{numerical-cal}
  M.~Kawasaki, K.~Nakayama and F.~Takahashi,
  ``Hilltop Non-Gaussianity,''
  JCAP {\bf 0901}, 026 (2009)
  [arXiv:0810.1585 [hep-ph]];
  M.~Kawasaki, T.~Kobayashi and F.~Takahashi,
  ``Non-Gaussianity from Axionic Curvaton,''
  arXiv:1210.6595 [astro-ph.CO];
 M.~Kawasaki, T.~Kobayashi and F.~Takahashi,
  ``Non-Gaussianity from Curvatons Revisited,''
  Phys.\ Rev.\ D {\bf 84} (2011) 123506
  [arXiv:1107.6011 [astro-ph.CO]].
\bibitem{Modulatedcurvatons}
  D.~Langlois and T.~Takahashi,
  ``Density Perturbations from Modulated Decay of the Curvaton,''
  arXiv:1301.3319 [astro-ph.CO];
  H.~Assadullahi, H.~Firouzjahi, M.~H.~Namjoo and D.~Wands,
  ``Modulated curvaton decay,''
  arXiv:1301.3439 [hep-th]
\bibitem{Modulatedcurvatons2}
S.~Enomoto, K.~Kohri and T.~Matsuda,
  ``Modulated decay in the multi-component Universe,''
  arXiv:1301.3787 [hep-ph];
  K.~Kohri, C.~-M.~Lin and T.~Matsuda,
  ``Delta-N Formalism for Curvaton with Modulated Decay,''
  arXiv:1303.2750 [hep-ph].

\bibitem{IR}
 G.~Dvali, A.~Gruzinov and M.~Zaldarriaga,
``A new mechanism for generating density perturbations from inflation,''
  Phys.\ Rev.\  D {\bf 69}, 023505 (2004)
  [arXiv:astro-ph/0303591].
\bibitem{IR-2}
  K.~Kohri, D.~H.~Lyth and C.~A.~Valenzuela-Toledo,
  ``Preheating and the non-gaussianity of the curvature perturbation,''
  JCAP {\bf 1002}, 023 (2010)
  [Erratum-ibid.\  {\bf 1009}, E01 (2011)]
  [arXiv:0904.0793 [hep-ph]].

\bibitem{end-of-inf}
 D.~H.~Lyth,
  ``Generating the curvature perturbation at the end of inflation,''
  JCAP\ {\bf 0511}, 006  (2005)
  [astro-ph/0510443];
  F.~Bernardeau, L.~Kofman and J.~-P.~Uzan,
  ``Modulated fluctuations from hybrid inflation,''
  Phys.\ Rev.\ D {\bf 70}, 083004 (2004)
  [astro-ph/0403315].

\bibitem{IPhaa}
  T.~Matsuda,
  ``Cosmological perturbations from an inhomogeneous phase transition,''
  Class.\ Quant.\ Grav.\  {\bf 26}, 145011 (2009).
  [arXiv:0902.4283 [hep-ph]];
 M.~Kawasaki, T.~Takahashi and S.~Yokoyama,
  ``Density Fluctuations in Thermal Inflation and Non-Gaussianity,''
  JCAP {\bf 0912}, 012 (2009)
  [arXiv:0910.3053 [hep-th]].

\bibitem{Lyth-gfc}
  D.~H.~Lyth,
   ``Can the curvaton paradigm accommodate a low inflation scale?,''
   Phys.\ Lett.\ B {\bf 579} (2004) 239
   [hep-th/0308110].


\bibitem{lm}
A.~D.~Linde and V.~F.~Mukhanov,
  ``Nongaussian isocurvature perturbations from inflation,''
  Phys.\ Rev.\  D {\bf 56} (1997) 535  [arXiv:astro-ph/9610219].


\bibitem{SVW}
  M.~Sasaki, J.~Valiviita and D.~Wands,
  ``Non-Gaussianity of the primordial perturbation in the curvaton model,''
  Phys.\ Rev.\ D {\bf 74}, 103003 (2006)
  [astro-ph/0607627].



\bibitem{Topological-curv}
 T.~Matsuda,
  ``Topological curvatons,''
  Phys.\ Rev.\ D {\bf 72}, 123508 (2005)
  [hep-ph/0509063].


\bibitem{Lyth-general}
  D.~H.~Lyth, K.~A.~Malik and M.~Sasaki,
  ``A General proof of the conservation of the curvature perturbation,''
  JCAP {\bf 0505}, 004 (2005)
  [astro-ph/0411220];



\bibitem{Langlois:2008vk} 
  D.~Langlois, F.~Vernizzi and D.~Wands,
  ``Non-linear isocurvature perturbations and non-Gaussianities,''
  JCAP {\bf 0812}, 004 (2008)
  [arXiv:0809.4646 [astro-ph]].


\bibitem{hilltop-matsuda}
  T.~Matsuda,
  ``Hilltop curvatons,''
  Phys.\ Lett.\ B {\bf 659}, 783 (2008)
  [arXiv:0712.2103 [hep-ph]].

\bibitem{fast-roll}
  A.~D.~Linde,
  ``Fast roll inflation,''
  JHEP {\bf 0111}, 052 (2001)
  [hep-th/0110195].

\bibitem{Lyth:2004nx} 
  D.~H.~Lyth and T.~Moroi,
  ``The Masses of weakly coupled scalar fields in the early universe,''
  JHEP {\bf 0405}, 004 (2004)
  [hep-ph/0402174].


\bibitem{iso-curv}
  M.~Beltran,
  ``Isocurvature, non-gaussianity and the curvaton model,''
  Phys.\ Rev.\ D {\bf 78}, 023530 (2008)
  [arXiv:0804.1097 [astro-ph]].

\bibitem{Curvatondynamics}
K.~Dimopoulos, G.~Lazarides, D.~Lyth and R.~Ruiz de Austri,\\
  ``Curvaton dynamics,''\\
  Phys.\ Rev.\ D {\bf 68}, 123515 (2003)[hep-ph/0308015].
\bibitem{Enq-Nu}
  K.~Enqvist and S.~Nurmi,\\
  ``Non-gaussianity in curvaton models with nearly quadratic potential,''\\
  JCAP {\bf 0510}, 013 (2005)[astro-ph/0508573].
\bibitem{Enq-LT}
K.~Enqvist, R.~N.~Lerner and O.~Taanila,\\
  ``Curvaton model completed,''\\
  JCAP {\bf 1112}, 016 (2011)[arXiv:1105.0498 [astro-ph.CO]].


\bibitem{hybcurv}
 K.~Dimopoulos, K.~Kohri and T.~Matsuda,
  ``The hybrid curvaton,''
  Phys.\ Rev.\ D {\bf 85}, 123541 (2012)
  [arXiv:1201.6037 [hep-ph]];
T.~Matsuda,
  ``Free light fields can change the predictions of hybrid inflation,''
  JCAP {\bf 1204}, 020 (2012)
  [arXiv:1204.0303 [hep-ph]].


\bibitem{IPre}
  T.~Matsuda,
  ``Cosmological perturbations from inhomogeneous preheating and
	multi-field trapping,''
  JHEP {\bf 0707}, 035 (2007)
  [arXiv:0707.0543 [hep-th]].


\bibitem{modulated-inflation}
  T.~Matsuda,
  ``Modulated Inflation,''
  Phys.\ Lett.\ B {\bf 665}, 338 (2008)
  [arXiv:0801.2648 [hep-ph]];
 T.~Matsuda,
  ``Modulated inflation from kinetic term,''
  JCAP {\bf 0805}, 022 (2008)
  [arXiv:0804.3268 [hep-th]].


\bibitem{matsuda-warm-curvaton}
  T.~Matsuda,
  ``Entropy production and curvature perturbation from dissipative curvatons,''
  JCAP {\bf 1009} (2010) 006
  [arXiv:1007.3636 [astro-ph.CO]];
  T.~Matsuda,
  ``Aspects of warm-flat directions,''
  Int.\ J.\ Mod.\ Phys.\ A {\bf 25}, 4221 (2010)
  [arXiv:0908.3059 [hep-ph]];
 T.~Matsuda,
  ``Evolution of the curvature perturbations during warm inflation,''
  JCAP {\bf 0906}, 002 (2009)
  [arXiv:0905.0308 [astro-ph.CO]].


\bibitem{N-flation-paper}
 S.~Enomoto and T.~Matsuda,
  ``Curvaton mechanism after multi-field inflation,''
  arXiv:1303.7023 [hep-ph];
  A.~R.~Liddle, A.~Mazumdar, F.~E.~Schunck,
  ``Assisted inflation,''
  Phys.\ Rev.\  {\bf D58}, 061301 (1998).
  [astro-ph/9804177];
  S.~Dimopoulos, S.~Kachru, J.~McGreevy, J.~G.~Wacker,
  ``N-flation,''
  JCAP {\bf 0808}, 003 (2008).
  [hep-th/0507205].
\end{thebibliography}
\end{document}